\documentclass{JHEP3}

\def \be {\begin{equation}}
\def \ee {\end{equation}}
\def \bea {\begin{eqnarray}}
\def \eea {\end{eqnarray}}
\def \nn {\nonumber}

\def \a {\alpha}
\def \b {\beta}
\def \g {\gamma}
\def \G {\Gamma}
\def \d {\delta}

\def \m {\mu}
\def \n {\nu}
\def \k {\kappa}

\def \s {\sigma}
\def \r {\rho}
\def \o {\omega}
\def \O {\Omega}
\def \th {\theta}
\def \Th {\Theta}

\def \t {\tau}
\def \dag {\dagger}
\def \p {\partial}

\def\bd{\begin{document}}
\def\ed{\end{document}}
\def\nn{\nonumber}
\def\bea{\begin{eqnarray}}
\def\eea{\end{eqnarray}}
\let\bm=\bibitem
\let\la=\label

\def\N{{\cal N}}
\def\sst{\scriptscriptstyle}
\def\thetabar{\bar\theta}
\def\Tr{{\rm Tr}}
\def\one{\mbox{1 \kern-.59em {\rm l}}}

\def\a{\alpha}      \def\da{{\dot\alpha}}
\def\b{\beta}       \def\db{{\dot\beta}}
\def\c{\gamma}  \def\C{\Gamma}  \def\cdt{\dot\gamma}
\def\d{\delta}  \def\D{\Delta}  \def\ddt{\dot\delta}
\def\e{\epsilon}        \def\vare{\varepsilon}
\def\f{\phi}    \def\F{\Phi}    \def\vvf{\f}
\def\h{\eta}
\def\k{\kappa}
\def\l{\lambda} \def\L{\Lambda}
\def\m{\mu} \def\n{\nu}
\def\o{\omega}
\def\P{\Pi}
\def\r{\rho}
\def\s{\sigma}  \def\S{\Sigma}
\def\t{\tau}
\def\th{\theta} \def\Th{\Theta} \def\vth{\vartheta}
\def\X{\Xeta}
\def\z{\zeta}
\def\w{\wedge}
\def\u{\underline}

\def\cA{{\cal A}} \def\cB{{\cal B}} \def\cC{{\cal C}}
\def\cD{{\cal D}} \def\cE{{\cal E}} \def\cF{{\cal F}}
\def\cG{{\cal G}} \def\cH{{\cal H}} \def\cI{{\cal I}}
\def\cJ{{\cal J}} \def\cK{{\cal K}} \def\cL{{\cal L}}
\def\cM{{\cal M}} \def\cN{{\cal N}} \def\cO{{\cal O}}
\def\cP{{\cal P}} \def\cQ{{\cal Q}} \def\cR{{\cal R}}
\def\cS{{\cal S}} \def\cT{{\cal T}} \def\cU{{\cal U}}
\def\cV{{\cal V}} \def\cW{{\cal W}} \def\cX{{\cal X}}
\def\cY{{\cal Y}} \def\cZ{{\cal Z}}

\def\ua{\underline{\alpha}}
\def\ub{\underline{\phantom{\alpha}}\!\!\!\beta}
\def\uc{\underline{\phantom{\alpha}}\!\!\!\gamma}
\def\um{\underline{\mu}}
\def\ud{\underline\delta}
\def\ue{\underline\epsilon}
\def\una{\underline a}\def\unA{\underline A}
\def\unb{\underline b}\def\unB{\underline B}
\def\unc{\underline c}\def\unC{\underline C}
\def\und{\underline d}\def\unD{\underline D}
\def\une{\underline e}\def\unE{\underline E}
\def\unf{\underline{\phantom{e}}\!\!\!\! f}\def\unF{\underline F}
\def\unm{\underline m}\def\unM{\underline M}
\def\unn{\underline n}\def\unN{\underline N}
\def\unp{\underline{\phantom{a}}\!\!\! p}\def\unP{\underline P}
\def\unq{\underline{\phantom{a}}\!\!\! q}
\def\unQ{\underline{\phantom{A}}\!\!\!\! Q}
\def\unH{\underline{H}}

\def\As {{A \hspace{-6.4pt} \slash}\;}
\def\bs {{b \hspace{-6.4pt} \slash}\;}
\def\Ds {{D \hspace{-6.4pt} \slash}\;}
\def\ds {{\del \hspace{-6.4pt} \slash}\;}
\def\ss {{\s \hspace{-6.4pt} \slash}\;}
\def\ks {{ k \hspace{-6.4pt} \slash}\;}
\def\ps {{p \hspace{-6.4pt} \slash}\;}
\def\pas {{{p_1} \hspace{-6.4pt} \slash}\;}
\def\pbs {{{p_2} \hspace{-6.4pt} \slash}\;}

\def\Fh{\hat{F}}
\def\Vh{\hat{V}}
\def\Xh{\hat{X}}
\def\ah{\hat{a}}
\def\xh{\hat{x}}
\def\yh{\hat{y}}
\def\ph{\hat{p}}
\def\xih{\hat{\xi}}

\def\psit{\tilde{\psi}}
\def\Psit{\tilde{\Psi}}
\def\tht{\tilde{\th}}

\def\At{\tilde{A}}
\def\Qt{\tilde{Q}}
\def\Rt{\tilde{R}}
\def\Nt{\tilde{N}}

\def\at{\tilde{a}}
\def\st{\tilde{s}}
\def\ft{\tilde{f}}
\def\pt{\tilde{p}}
\def\qt{\tilde{q}}
\def\vt{\tilde{v}}
\def\nt{\tilde{n}}

\def\delb{\bar{\partial}}
\def\bz{\bar{z}}
\def\bD{\bar{D}}
\def\bB{\bar{B}}

\def\bk{{\bf k}}
\def\bl{{\bf l}}
\def\bp{{\bf p}}
\def\bq{{\bf q}}
\def\br{{\bf r}}
\def\bx{{\bf x}}
\def\by{{\bf y}}
\def\bR{{\bf R}}
\def\bV{{\bf V}}

\def\d{\delta}\def\D{\Delta}\def\ddt{\dot\delta}

\def\p{\partial} \def\del{\partial}
\def\xx{\times}
\def\uno{\mbox{1 \kern-.59em {\rm l}}}

\def\trp{^{\top}}
\def\inv{^{-1}}
\def\dag{{^{\dagger}}}
\def\pr{^{\prime}}

\def\rar{\rightarrow}
\def\lar{\leftarrow}
\def\lrar{\leftrightarrow}

\title{M5-branes and Wilson Surfaces  }
\author{Bin Chen\footnote{On visit of the Abdus Salam International Center for Theoretical Physics}\\
Department of Physics,\\
Peking University,\\
Beijing 100871, P.R. China\\
\email{bchen01@pku.edu.cn}}

\author{Wei He\\
Institute of Theoretical Physics\\
 Chinese Academy of Science,\\
 Beijing 100080, P.R. China\\
\email{weihe@itp.ac.cn}}

\author{Jun-Bao Wu\\International School for Advanced Studies (SISSA)\\
via Beirut 2-4, I-34014 Trieste, Italy\\
\email{wujunbao@sissa.it}}

\author{Liang Zhang \\
Department of Physics,\\
Peking University,\\
Beijing 100871, P.R. China\\
 \email{lzhang021@gms.phy.pku.edu.cn}}

\date{\today}

\abstract{We investigate the M5-brane description of the Wilson
surface operators in six-dimensional (2,0) superconformal field
theory from AdS/CFT correspondence. We consider the Wilson surface
operators in high-dimensional representation, whose description
could be M5-brane string soliton solutions in $AdS_7\times S^4$
background. We construct such string soliton solutions from the
covariant M5-brane equations of motion and discuss their
properties. The supersymmetry analysis shows that these solutions
are half-BPS. We also discuss the subtle issue on the boundary
terms.}

\preprint{\ SISSA-48/2007/EP}
\newpage

\begin{document}

\section{Introduction}

One of the most important achievements in string theory is AdS/CFT
correspondence\cite{Mal97}, which states that the quantum gravity
in Anti-de-Sitter(AdS) spacetime is dual to the large N limit of
the superconformal field theory at AdS boundary. It not only opens
a new window to study the quantum gravity from its dual field
theory, but also supplies a new tool to study the gauge theory
from its gravity dual.

The best studied case in AdS/CFT correspondence is $AdS_5/SYM_4$,
where the boundary field theory is an $\cN=4$ super-Yang-Mills
gauge theory. Many works have been devoted to the study of the
correspondence in this case. Among them, the study of the
supersymmetric Wilson loop operators\cite{Mal98,Gross} is one of
the most interesting issues. The Wilson loop operators in gauge
theory is defined to be the holonomy of the gauge field around a
contour. Their expectation values characterize the phase of the
theory. From AdS/CFT dictionary, the expectation values could be
calculated by considering a fundamental string ending on the
boundary of $AdS$ along the path specified by the Wilson loop
operator. The area of the string worldsheet bounded by the loop
may give the expectation value of the Wilson loop operator, after
appropriate regularization. However, this is not the whole story.
In the field theory side, it has been found that the calculation
of the expectation value of 1/2-BPS circular Wilson loop in
$SYM_4$ could be reduced to a corresponding calculation in a
zero-dimensional matrix model\cite{Zarembo}. There are two
remarkable things. One is that though the expectation value of the
straight half-BPS Wilson line is exactly one, the one of the
circular loop is more involved, being a function of the 't Hooft
coupling. The circular loop is related to the straight line by
conformal transformation. The difference of the expectation values
of two cases  comes from the conformal anomaly of the boundary.
The other remarkable point is that the calculation through the
matrix model gives the expectation value not only to all orders of
't Hooft coupling $\lambda=g^2_{YM}N$ but also to all orders of
$1/N$\cite{Drukker01}, which is just the string coupling $g_s$
from the correspondence. This indicates that the calculation in
gravity side should go beyond the free string limit.

Inspired by the field theory result,  it has been found that the
all-genus expectation value of the 1/2 BPS Wilson loop operators
is better described by the dynamics of the D3-brane\cite{Drukker}
or D5-brane\cite{Yamaguchi:2006D5}, not just by the minimal
surface of the string world-sheet\cite{Gross}, especially when the
charges of the string are large. Simply speaking, for the Wilson
loop in the symmetric representation or the multiply wound Wilson
loop, the many coincident fundamental strings could interact
themselves and be described in terms of the dynamics of D3-branes
with electric flux\cite{Drukker}. This is reminiscent of the giant
gravitons\cite{McGreevy00, Grisaru00, Hashimoto00}. For the Wilson
loops in anti-symmetric representation, the interaction of the
strings leads to a better description in terms of the dynamics of
the D5-brane with electric flux\cite{Yamaguchi:2006D5}. One may
also understand the above picture from Myers
effect\cite{Myers:1999}: the string worldsheet in the five-form
field strength background may blow up in the transverse directions
to form dielectric brane\cite{Rodriguez2006}. The generalization
to the Wilson-'t Hooft operators could be found in \cite{ChenHe}.

Motivated by the success in the Wilson loop case, we are led to
consider its cousin in 6-dimensional field theory. In this case,
we have $AdS_7/CFT_6$ correspondence\cite{AdS7CFT}, where the
$CFT_6$ is a six-dimensional superconformal field theory. This
duality originate from the the description of M5-brane in
11-dimensional M-theory. The near horizon limit of M5-brane
gravity solution in 11-dimensional supergravity is $AdS_7 \times
S_4$. And the low energy effective field theory of coincident
M5-brane is a $(2,0)$-superconformal field
theory\cite{Andy95,Witten95}. This field theory is quite
mysterious\cite{Seiberg}: its field content is of a tensor
multiplet, including a 2-form $B_{\mu\nu}$, four fermions and five
scalars; the field strength of 2-form is (anti)self-dual. The
existence of the 2-form gauge field implies that there exist
string like excitations in the theory. However, there is no
lagrangian formulation of the chiral 2-form, even though the
chiral theory is still a local interacting field
theory\cite{Witten96}. The theory has been suggested to be
described by DLCQ matrix theory\cite{Seiberg97}. In any sense, it
has not been well understood. The AdS/CFT correspondence supplies
a new tool to probe this nontrivial six-dimensional field theory.
The weak version of the correspondence says that the large $N$
limit of the $(2,0)$ field theory is dual to 11D supergravity on
$AdS_7 \times S^4$\cite{Mal97}. Some properties of the field
theory has been studied from its dual gravity, including the
correlation functions of the chiral primary operators and the
Wilson surface operators\cite{Mal98,Corrado}. The Wilson surface
operators in $(2,0)$ theory  could be formally defined
as\cite{Ganor}
 \be
 W_0(\Sigma)=\exp i\int_\Sigma B^+,
 \ee
where $\Sigma$ is a two-dimensional surface. From AdS/CFT
correspondence, its expectation value could be calculated from the
membrane action as
 \be\label{WS}
 <W_0(\Sigma)>=e^{-S}
 \ee
 where $S$ is the action of the membrane whose worldvolume boundary
 is $\Sigma$. The action is divergent and
needs renormalization. Unlike its cousin the Wilson loop, the
surface operator has conformal anomaly since even dimensional
submanifold observable is conformally anomalous\cite{Witten99}.
For example, the membrane action corresponding to the spherical
Wilson surface has both quadratic and logarithmic divergences. The
existence of logarithmic divergence indicates that the expectation
value of Wilson surface may not be well-defined. Unlike the Wilson
loop operator, only abelian Wilson surface operator has been
discussed in the literature. The field theory study could be found
in \cite{Gustavsson}: the conformal anomaly of abelian Wilson
surface operator was calculated in $A_1$ field theory. It is very
hard to consider the nonabelin Wilson surface in the field theory.
In \cite{Corrado}, from AdS/CFT correspondence, the Wilson surface
operators in the fundamental representation has been studied. It
would be interesting to see if we can find the M5-brane
description of the Wilson surface in higher dimensional
representation, just like the case in the Wilson loop. Especially,
we want to consider the 1/2-BPS Wilson surface operators. We will
show in this paper that this is feasible.

We must be cautious in talking about the BPS Wilson surface in
higher dimensional representation, since we have no idea on how to
define it rigorously in the field theory. Formally we can define
the Wilson surface in representation $R$ to be
 \be
 W_R(\Sigma)=\Tr_R P[\exp i\int_\Sigma (B^++\cdots)],
 \ee
 where $\cdots$ denotes the possible scalar fields and fermions so
 that we have a 1/2-BPS Wilson surface operators. Without a
 lagrangian formulation of the chiral 2-form field theory, it is
 not clear how to find the half-supersymmetric operators and what
 the expectation value of $W_R(\sigma)$ really means. It could be
 better to understand the situation from brane picture. The
 surface $\Sigma$ could be taken as the intersection of membrane
 with the M5-branes. In brane configuration, a Wilson surface operator in  the rank $k$ symmetric representation,
 corresponds to a $k$-wound membrane ending on $\Sigma$\footnote{Strictly speaking,
 there could be a small difference between symmetric representation and multi-wound surfaces.
 In the Wilson loop case, this issue has been addressed in
 \cite{Yamaguchi0612}.}, while
 a Wilson surface operator in the rank $m$ antisymmetric
 representation corresponds to $m$ membranes ending on
 M5-branes. These are the two most simplest cases, which we
 will study in this paper. For more general representation, one
 may get the brane picture from the lesson in the Wilson loop
 case\cite{Gomis06}.

From $AdS_7/CFT_6$ correspondence, the expecation value of the
Wilson surface could be calculated by the regularized volume of
the membrane ending on the M5-brane with boundary $\Sigma$. For
the multi-wrapped Wilson surface, the interaction among membranes
may induce a blow-up of the membrane to M5-brane wrapping $S^3$.
The dynamics of M5-brane with self-dual field strength encodes the
information of the Wilson surface. Analogue to the Wilson line
case, the expectation value of the Wilson surfaces could still be
calculated from (\ref{WS}) but now the action is the M5-brane
action with appropriate boundary terms. However, unlike the
Dp-brane, the M5-brane dynamics is much more involved. From the
M5-brane point of view, the membranes are the self-dual string
soliton. The first step in our investigation is to find the
self-dual string soliton solution of M5-brane, whose worldvolume
is embedded in $AdS_7\times S^4$ background. In a curved
spacetime, the equations of motion of M5-brane looks forbidding
and hard to solve. There is no careful discussion on this issue.
Most of the discussion on M5-brane string soliton have been
focused on the flat spacetime. We will start from the covariant
equations of motion and construct the M5-brane soliton solutions
corresponding to the higher-dimensional Wilson surfaces. The
similar brane configurations has been discussed in \cite{Lunin07}
in Pasti-Sorokin-Tonin(PST) formalism. Our results are in
agreement with the ones in \cite{Lunin07}.

This investigation is rewarding. It may help us to understand
better the dynamics of the M5-brane, give prediction on the
expectation value of multi-wrapped Wilson surface, which has not
been worked out in six-dimensional (2,0) theory. It could also
shed light on the membrane interaction and the possible Myers
effect in M-theory.

The paper is organized as follows. In the next section, we give a
brief review of the M5-brane equations of motions. In section 3,
we work out M5-brane string soliton solutions, whose worldvolume
is embedded into $AdS_7$ and is of topology $AdS_3\times S^3$. We
consider both the straight Wilson surface and the spherical Wilson
surface. These soliton solutions describes the Wilson surface
operator in the symmetric representation. In section 4, we study
another kind of  string soliton solutions with the same topology
but with $S^3$ part in $S^4$. They describe the straight and
spherical Wilson surface operators in the antisymmetric
representation. For both kinds of solutions, we discuss their
properties, including charges, bulk actions, boundary terms. In
section 5, we show that our solutions are half-BPS. This suggests
that the corresponding Wilson surface operators are also 1/2-BPS
in the field theory.  We end with the conclusion and discussions.
In Appendix, we list various connections used in our calculation.

\section{The M5-brane equations of motion and actions}

In this section, we give a brief review of the M5-brane covariant
equations of motions in curved spacetime. For a review on various
aspects of M5-brane, see \cite{Sezgin99}.

The M5-brane covariant equations of motion in eleven dimension was
first proposed in \cite{Sezgin97} in superembedding
formalism\cite{Howe96}, and was then rederived by requiring
$\k$-symmetry of an open M2-brane ending on the
M5-brane\cite{Chu97}. For other derivations from various actions,
please see \cite{Sundell97,Sorokin97}. We only care about the
bosonic components of the equations, which include the scalar
equation and tensor equation. The scalar equation takes the form
 \be\label{scalareq}
 G^{mn}\nabla_m \cE_n^{\underline
 c}=\frac{Q}{\sqrt{-g}}\epsilon^{m_1\cdots
 m_6}\big(\frac{1}{6!}H^{\underline a}_{~m_1\cdots
 m_6}+\frac{1}{(3!)^2}H^{\underline
 a}_{~m_1m_2m_3}H_{m_4m_5m_6}\big)P_{\underline
 a}^{~\underline c}
 \ee
 and the tensor equation is of the form
 \be\label{tensoreq}
 G^{mn}\nabla_mH_{npq}=Q^{-1}(4Y-2(mY+Ym)+mYm)_{pq}.
 \ee
 Here our notation is as follows: indices from the
beginning(middle) of the alphabet refer to frame(coordinate)
indices, and the underlined indices refer to target space ones.

Let us spend some time to explain the quantities in the above
equation. There exists a self-dual 3-form field strength $h_{mnp}$
on the M5-brane worldvolume. From it, we can define
 \bea
 k_m^{~n}&=&h_{mpq}h^{npq}, \\
 Q&=&1-\frac{2}{3}\Tr k^2, \\
 m_p^{~q}&=&\delta_p^{~q}-2k_p^{~q}, \\
 H_{mnp}&=&4Q^{-1}(1+2k)_m^{~q}h_{qnp}
 \eea
 Note that $h_{mnp}$ is self-dual with respect to worldvolume
 metric but not $H_{mnp}$. The
 induced metric is simply
 \be
 g_{mn}=\cE_m^{\underline a}\cE_n^{\underline b}\eta_{{\underline
 a}{\u b}}
 \ee
 where
 \be
 \cE_m^{\underline a}=\p_mz^{\underline m}E_{\underline m}^{\underline
 a}.
 \ee
Here $z^{\underline m}$ is the target spacetime coordinate, which
is a function of worldvolume coordinate $\xi$ through embedding, and
$E_{\underline m}^{\underline
 a}$ is the component of target space vielbein. From the induced
 metric, we can define another tensor
 \be
 G^{mn}=(1+\frac{2}{3}k^2)g^{mn}-4k^{mn}.
 \ee
 We also have
 \be
 P_{\underline a}^{~\underline c}=\delta^{\underline
 c}_{\underline a}-\cE_{\underline a}^m\cE_m^{~{\underline c}}.
 \ee

Note that in the scalar equation of motion, the covariant
derivative $\nabla_m\cE_n^{\underline c}$ involves not only the
Levi-Civita connection of the M5-brane worldvolume but also the
spin connections of the target spacetime geometry. More precisely
one has
 \be
 \nabla_m\cE_n^{\underline c}=\p_m\cE_n^{\underline c}-\G^p_{mn}\cE_p^{\underline
 c}+\cE_m^{\underline a}\cE_n^{\underline b}\o^{\underline c}_{{\underline a}{\underline
 b}}
 \ee
 where $\G^p_{mn}$ is the Christoffel symbol with respect to the induced worldvolume
 metric and $\o^{\underline c}_{{\underline a}{\underline
 b}}$ is the spin connection of the background spacetime.

 Moreover, there is a 4-form field strength $H_{{\underline a}_1\cdots {\underline a}_4}$ and its Hodge dual
 7-form field strength $H_{{\underline a}_1\cdots {\underline
 a}_7}$:
 \bea
 H_4&=&dC_3 \nn\\
 H_7&=&dC_6+\frac{1}{2}C_3\w H_4
 \eea
 The frame indices on $H_4$ and $H_7$ in the above equations (\ref{scalareq},\ref{tensoreq}) have
 been converted to worldvolume indices with factors of
 $\cE_m^{\underline c}$.
  From them, we can define
 \be
 Y_{mn}=[4\star {\underline H}-2(m\star {\underline H}+\star {\underline H}m)+m\star {\underline H}m]_{mn},
 \ee
where \be \star {\underline
H}^{mn}=\frac{1}{4!\sqrt{-g}}\epsilon^{mnpqrs}{\underline
H}_{pqrs} \ee

The field $H_{mnp}$ is defined by
 \be
 H_3=dA_2-{\u C}_3,
 \ee
 where $A_2$ is a 2-form gauge potential and ${\u C}_3$ is the pull-back
 of the bulk gauge potential. From its definition, $H_3$ satisfies the
Bianchi
 identity
 \be
 dH_3=-{\underline H}_4
 \ee
 where ${\underline H}_4$ is the pull-back of the target space 4-form flux.
 Note that different from the 3-form field $h_3$ which is self-dual on the worldvolume
 of M5-brane, $H_3$ satisfies a nonlinear self-duality condition:
 \be
 \ast H_{mnp}=Q^{-1}G_m^{~q}H_{npq}.
 \ee
This condition could be rewritten in the following form
 \be\label{selfdual}
 \ast H_3=\frac{\p {\cal K}}{\p H_3},
 \ee
 where
 \be
 {\cal
 K}=2\sqrt{1+\frac{1}{12}H^2+\frac{1}{288}(H^2)^2-\frac{1}{96}H_{abc}H^{bcd}H_{def}H^{efa}}.
 \ee

It would be nice to derive the above equations of motion from an
action. However, compared to the Dp-brane action, which is just a
Dirac-Born-Infeld(DBI)-type action, M5-brane action is much more
subtle since it describe a self-interacting chiral 2-form whose
field strength is self-dual.
 In
\cite{Sorokin9701,Sorokin9711}, a manifestly supercovariant and
kappa-invariant action has been constructed. It contains an
auxiliary scalar, from which the self-duality condition could be
derived as an equation of motion. The covariant bosonic action is
of a DBI-like form\footnote{For an equivalent formulation with the
same philosophy, see\cite{Schwarz97}.}
 \bea\label{covariant}
 S_c=T_5\int
 d^6x\left(\sqrt{-\det(g_{mn}+i{\tilde H}_{mn})}-\frac{\sqrt{-g}}{4}{\tilde
 H}^{mn}H_{mn}\right)-T_5\int Z_6
 \eea
 where
 \bea
 Z_6&=&\underline{C}_6-\frac{1}{2}\underline{C}_3\w H_3,
 \eea
and $T_5$ is the tension of the M5-brane:
 \be
 T_5=\frac{1}{(2\pi)^5l^6_p}.
 \ee
 Here $Z_6$ is the
Wess-Zumino term, in which ${\underline C}_6$ and ${\underline
C}_3$ are the pull-back of the target space gauge potential. In
the action, one has
 \bea
 {\tilde H}^{mn}&=&(\ast H)^{mnp}v_p,\\
 H_{mn}&=&H_{mnp}v^p,
 \eea
 by introducing an auxiliary field $b$ whose normalized derivative is
  \be
  v_p=\frac{\p_p b}{\sqrt{-g^{mn}\p_m b\p_n b}}.
  \ee
  Note that the vector $\vec{v}$ is timelike, $v_pv^p=-1$ and
  one has the freedom in choosing $v_p$. The equation of motion of
  the auxiliary field $b$ is not independent. It is a consequence
  of the equation of motion of the 2-form gauge potential, which
  takes the following form after appropriate gauge fixing:
   \be\label{sdual}
   H_{mn}={\cal V}_{mn},
   \ee
   where
   \be
   {\cal V}_{mn}=-\frac{2}{\sqrt{-g}}\frac{\d\sqrt{-\det(g_{mn}+i{\tilde
   H}_{mn})}}{\d {\tilde H}^{mn}}.
   \ee
The relation (\ref{sdual}) is actually a generalized self-dual
condition.

This proposal has some troubles in defining a proper partition
function, since the topological class of auxiliary scalar would
break some symmetries of M-theory, as pointed out in
\cite{Witten96}. The resolution of this problem is to embed the
chiral theory into a non-chiral one. In \cite{Sundell97}, a
nonchiral M5-brane action for unconstrained 2-form gauge potential
has been constructed. In this action, one has to impose a
non-linear self-duality condition to ensure kappa-symmetry. And
the equation of motion for 2-form potential is equivalent to the
Bianchi identity. The action is given by
 \be\label{nc}
 S=S_{M5}-S_{WZ}=T_5\int (\frac{1}{2}\star {\cal K}-Z_6)
 \ee

There are two remarkable relation on ${\cal K}$, when the
nonlinear self-duality condition (\ref{selfdual}) holds:
 \bea
  {\cal K}&=&2K=2\sqrt{1+\frac{1}{24}H^{mnp}H_{mnp}},\nn\\
  K&=&2Q^{-1}-1.
 \eea

It has been proved in \cite{Sorokin97} that the two different
actions (\ref{covariant}) and (\ref{nc}) are equivalent, leading
to the same set of equations of motion.

\section{M5-brane description of the Wilson surface in the symmetric
representation}

The $AdS_7\times S^4$ spacetime could be taken as the near horizon
geometry of M5-brane gravity solution. It is also the maximally
supersymmetric solution in 11-dimensional supergravity, whose
equations of motion take the form,
 \bea
 R_{MN}&=&\frac{1}{2\xx 3!}H_{MPQR}H_{N}^{~PQR}-\frac{1}{6\xx
 4!}g_{MN}H_{PQRS}H^{PQRS}, \\
 0&=&\p_M\sqrt{-g}H^{MNPQ}+\frac{1}{2\xx
 (4!)^2}\epsilon^{M_1\cdots M_8NPQ}H_{M_1\cdots M_4}H_{M_5\cdots
 M_8}\eea
 where the indices take values from $0$ to $10$.
 The metric and background 4-form flux of $AdS_7\times S^4$ are
 \bea
 ds^2&=&\frac{R^2}{y^2}(dy^2-dt^2+dx^2+dr^2+r^2d\Omega_3^2)+\frac{R^2}{4}d\Omega_4^2\nn\\
 H_4&=&\frac{3R^3}{8}\sin^3\z_1\sin^2\z_2\sin\z_3d\z_1\w d\z_2\w
 d\z_3\w d\z_4
 \eea
 where $d\O_3^2$ is the metric of $S^3$ and $d\O_4^2$ is the
 metric of $S^4$. The 4-form field strength fills in $S^4$, in
 which $\z_i, i=1,2,3$ are three angular coordinates in $S^4$.
 Note that the radius of $AdS_7$ is twice the one of $S^4$.
 Here since our discussion following will focus on the $AdS_7$ in
 this section, we rescale its radius to be $R$. From the
 $AdS_7/CFT_6$ duality, we know that
 \be
 R=(8\pi N)^{\frac{1}{3}}l_p,
 \ee
 where $l_p$ is the Planck constant in eleven dimensions.

\subsection{Straight Wilson surface}

The standard description of the Wilson surface in AdS/CFT
correspondence is the boundary of a minimal area membrane
worldvolume in $AdS_7\times S^4$ background. For the straight
Wilson surface, we can set the membrane worldvolume to be extended
in the directions $y,t,x$ and fixed at $r=0$. From the AdS/CFT
dictionary, the expectation value of the Wilson surface operator
depends on the volume of the membrane through (\ref{WS}). In the
straight Wilson surface case, the volume with respect to the
induced metric is
 \bea
 V_2&=&\int \frac{R^3}{y^3}dy dt dx\nn\\
 &=& TX\frac{R^3}{2y^2_0},
 \eea
where $T,X$ is the lengths of the $t,x$ directions. Here we have
introduced a cutoff $y_0$ to regularize the volume. The action is
just
 \be\label{strmembrane}
 S=T_2 V_2=\frac{N}{\pi}TX\frac{1}{y_0^2},
 \ee
where $T_2=\frac{1}{(2\pi)^2l^3_p}$ is the membrane tension. The
action is proportional to the area of the surface and is also of
quadratic divergence.

 To have a M5-brane description of the Wilson surface in the high rank
 representation, we have
 to find the appropriate M5-brane solution first. Inspired by the
 study of the Wilson loop, one may attempt to try a M5-brane with
 worldvolume $AdS_3\times S^3$\cite{Berman01}. The induced membrane worldvolume
 is an $AdS_3$ and the blow-up of the background flux gives an
 $S^3$. In the case of  the straight Wilson surface, let
  the worldvolume coordinates of M5-brane
be $\xi_i, i=0,\cdots 5$ and the embedding be
 \bea
 \xi_0=t,~~~ \xi_1=x,~~~ \xi_2=r,~~~ y=f(r), \\
 \xi_3=\a,~~~ \xi_4=\b, ~~~ \xi_5=\g
 \eea
 where $\a,\b,\g$ are the angular coordinates of $S^3$. The
 induced metric is then
 \bea
 ds^2_{\mbox{ind}}&=&\frac{R^2}{f^2}(-d\xi_0^2+d\xi_1^2+(1+f^{\pr
 2})d\xi_2^2+r^2d\O_3^2)\nn\\
 &=&\frac{R^2}{f^2}(-dt^2+dx^2+(1+f^{\pr
 2})dr^2)+\frac{R^2r^2}{f^2}(d\a^2+\sin^2\a d\b^2+\sin^2\a
 \sin^2\b d\g^2)\nn\\
 & &\label{indmetric}
 \eea
 where the prime denotes the derivative with respect to $r$. Without causing confusion, we
 simply let $t,x,r,\a,\b,\g$ be the
 coordinates of the M5-brane worldvolume.

There is a self-dual 3-form field strength in the M5-brane
worldvolume. Let us assume it to be
 \be
 h_3=\frac{a}{2}(1+\star_{\mbox{ind}})\sqrt{\det G}d\a \wedge d\b
 \wedge d\g
 \ee
 where $a$ could be a function of $r$ and $\det G$ is the determinant of the metric of $S^3$. In our
 case, we have
 \be
 h_3=\frac{a}{2}(\frac{R}{f})^3(r^3\sin^2\a\sin\b d\a\w d\b\w
 d\g+\sqrt{1+f^{\pr 2}}dt\w dx\w dr).
 \ee

Then we can calculate the relevant quantities $k^{mn}, G^{mn}$
etc.. Here we list the quantities which will be useful to our
following discussion:
 \bea
 k^2&=&k_{mn}k^{mn}=\frac{3}{2}a^4,\nn\\
 k_m^{~n}&=&\left(\begin{array}{cc}
 -\frac{a^2}{2}I_3&0\\
 0&\frac{a^2}{2}I_3
 \end{array}\right),\nn\\
  G^{tt}&=&-G^{xx}=-(\frac{f}{R})^2(1+a^2)^2,\nn\\
 G^{rr}&=&(\frac{f}{R})^2\frac{(1+a^2)^2}{1+f^{\pr 2}}, \nn\\
 G^{\a\a}&=&(\frac{f}{Rr})^2(1-a^2)^2,\nn\\
 G^{\b\b}&=&(\frac{f}{Rr\sin\a})^2(1-a^2)^2~~~\nn\\
  G^{\g\g}&=&(\frac{f}{Rr\sin\a\sin\b})^2(1-a^2)^2, \label{kG}
 \eea
 where $I_3$ is a rank 3 identity matrix, and
 \be
 H_3=2a(\frac{R}{f})^3(\frac{\sqrt{1+f^{\pr
 2}}}{1+a^2}dt\w dx \w dr +\frac{r^3}{1-a^2}\sin^2\a\sin\b d\a \w
 d\b\w d\g)
 \ee

The non-chiral five-brane action is just
 \be
 S=T_5\int dtdxdrd\a d\b d\g (\frac{R}{f})^6r^3\sin^2\a\sin\b
 (\sqrt{1+f^{\pr 2}}\frac{1+a^4}{1-a^4}-1)
 \ee

  Since there is no pull-back of bulk 4-form field strength on the
 M5-brane worldvolume, we have $dH_3=0$, which gives the
 constraint
 \be
 \frac{a}{1-a^2}\frac{r^3}{f^3}=\mbox{constant}
 \ee

The equation of motion on the tensor $H_{npq}$, in this case, is
 \be
 G^{mn}\nabla_m H_{npq}=0.
 \ee
Here $\nabla_m$ is the covariant derivative with respect to the
induced metric. We list the detailed Levi-Civita connection in
Appendix. It is straightforward to check that the above equation
is satisfied, provided that $a$ is a constant. Then from $dH=0$,
we can determine \be f(r)=\kappa r,\ee where $\kappa$ is just a
constant. This is reminiscent of the solution in the straight
Wilson line case\cite{Drukker,ChenHe} in $AdS_5\times S^5$ and the
spiky solution in flat spacetime\cite{Callan97,Townsend97}. Then
the induced  metric of M5-brane is
 \be
 ds^2=\frac{R^2}{\k^2r^2}(-dt^2+dx^2+(1+\k^2)dr^2)+\frac{R^2}{\k^2}(d\a^2+\sin^2\a d\b^2+\sin^2\a
 \sin^2\b d\g^2). \label{indmetric01}
 \ee
 This indicates that the worldvolume of M5 brane is actually
 $AdS_3\times S^3$, with radius $\frac{R\sqrt{1+\k^2}}{\k}$ in
 $AdS_3$ and radius $\frac{R}{\k}$ in $S^3$.
 The self-dual 3-form field strength is then
 \be
 h_3=\frac{a}{2}\frac{R^3}{\k^3}(\sin^2\a\sin\b d\a\w d\b\w
 d\g+\frac{\sqrt{1+\k^2}}{r^3}dt\w dx \w dr)
 \ee
 and
 \be
 H_3=2a\frac{R^3}{\k^3}(\frac{1}{1-a^2}\sin^2\a\sin\b d\a\w d\b\w
 d\g+\frac{\sqrt{1+\k^2}}{(1+a^2)r^3}dt\w dx \w dr)
 \ee

For the scalar equation of motion, it is more involved. In our
case, we have
 \bea
 \cE^{\underline 0}_t=\frac{R}{\k r}, ~~~ \cE^{\underline 1}_{r}=\frac{R}{r}, ~~~\cE^{\underline
 2}_{x}=\frac{R}{\k r}, ~~~ \cE^{\underline 3}_{r}=\frac{R}{\k r},\nn\\
 \cE^{\underline 4}_{\a}=\frac{R}{\k},~~~ \cE^{\underline
 5}_{\b}=\frac{R\sin\a}{\k}, ~~~ \cE^{\underline
 6}_{\g}=\frac{R\sin\a\sin\b}{\k},
 \eea
where we have set the veilbein of $AdS_7$ part of the target
spacetime as
 \bea
 \hat{\th}^0=\frac{R}{y}dt,~~~\hat{\th}^1=\frac{R}{y}dy,~~~\hat{\th}^2=\frac{R}{y}dx,~~~\hat{\th}^3=\frac{R}{y}dr,\nn\\
 \hat{\th}^4=\frac{Rr}{y}d\a,~~~\hat{\th}^5=\frac{Rr\sin\a}{y}d\b,~~~\hat{\th}^6=\frac{Rr\sin\a\sin\b}{y}d\g.
 \eea
The corresponding spin connection could be found in Appendix.

 The scalar equation of motion involves
 \be
 \nabla_m\cE_n^{\underline c}=\p_m\cE_n^{\underline c}-\G^p_{mn}\cE_p^{\underline
 c}+\cE_m^{\underline a}\cE_n^{\underline b}\o^{\underline c}_{{\underline a}{\underline
 b}}
 \ee
 where $\G^p_{mn}$ is the Christoffel symbol and $\o^{\underline c}_{{\underline a}{\underline
 b}}$ is the spin connection of the background spacetime. The
 calculation shows that
 \be
 G^{mn}\nabla_m\cE_n^{\underline c}=0,\hspace{5ex} \mbox{except ${\underline
 c}={\underline 1}$ or ${\underline 3}$}. \ee
 The nontrivial components come from ${\underline c}={\underline 1}$ or ${\underline
 3}$. The right hand side of the scalar equation of motion consists of
 the matrix $P^{\underline c}_{\underline a}=\delta^{\underline
 c}_{\underline a}-\cE_{\underline a}^m\cE_m^{~{\underline c}}$,
 which has nonvanishing components
  \be
 P^{~\underline c}_{\underline a}=\left(\begin{array}{cc}
 \frac{1}{1+\k^2}&-\frac{\k}{1+\k^2}\\
 -\frac{\k}{1+\k^2}&\frac{\k^2}{1+\k^2}
 \end{array}\right).
 \ee
 where ${\underline a},{\underline c}$ take values ${\underline 1},{\underline
 3}$.

 For the background flux, we have a dual 7-form field strength in
 $AdS_7$ part,
  \be
  H_{{\underline 0}{\underline 1}\cdots {\underline
  6}}=\frac{6}{R}
  \ee
Note that our convention is a little different from the literature
by a factor $2$ since we have rescaled the radius of $AdS_7$.

On the right hand side of the scalar equation, only 7-form field
strength contributes since the M5-brane worldvolume is embedded
simply into $AdS_7$ and there is no induced 4-form field strength
on it.

 It turns out that the nontrivial components ${\underline c}={\underline 1}$ and ${\underline
 3}$ of the scalar equation of motion give the same constraint:
  \be
  \frac{(1+a^2)^2}{1+\k^2}+(1-a^2)^2=-2\frac{1-a^4}{\sqrt{1+\k^2}}.
 \ee
 This equation could be solved and gives the relation
  \be\label{ak}
  a=\frac{\pm\k}{\sqrt{1+\k^2}-1}
  \ee

In short, we have obtained a  string soliton solution of M5-brane,
once the relation (\ref{ak}) is satisfied. We will show in section
5 that our solution is indeed half-BPS. This solution matches with
the one found in section 2.3 in \cite{Lunin07}.

Let us consider some properties of this string soliton solution.
One can calculate the charges of this string soliton. Since our
solution could be taken as M2-branes ending on M5-brane, with
M2-brane worldvolume extending along $t,x,r$, the charges could be
calculated by
 \bea
 Q_E&=&\frac{1}{{\mbox{Vol}(S^3)}}\int_{S^3} \star H \label{QE}\\
 Q_M&=&\frac{1}{{\mbox{Vol}(S^3)}}\int_{S^3}  H \label{charges}
 \eea
 where $S^3$ is the transverse $S^3$ and $\star$ here means the Hodge dual
 with respect to the metric of the M5-brane worldvolume without the string
 soliton. Here is a subtlety. If we take the strategy suggested in \cite{Howe96}
 and think that
 the metric of the 5-brane worldvolume without string soliton is just
 \be
 ds^2=\frac{R^2}{\k^2r^2}(-dt^2+dx^2+dr^2)+\frac{R^2}{\k^2}(d\a^2+\sin^2\a d\b^2+\sin^2\a
 \sin^2\b d\g^2)
 \ee
 which indicates that the worldvolume is a $AdS_3\times S^3$ with the same
 radius, then our solution  has opposite electric and magnetic charge:
 \bea
 Q_E&=&\pm\frac{R^3}{l^3_p\k^2} \\
 Q_M&=&\mp\frac{R^3}{l^3_p\k^2}.
 \eea
However, the above treatment could be problematic. When we turn
off the charge, the $S^3$ part shrinks also so we have no M5-brane
worldvolume anymore. This means that the solution is not the same
self-dual string soliton on M5-brane as the one discussed in
\cite{Howe96}. It is more like the case in \cite{Drukker}, where a
D3-brane is blown up by the Wilson line. Analogously, it would be
better to calculate the charge from the action itself. For the
magnetic charge, the above result is fine. But for the electric
charge, it could be better to start from the conjugate momentum of
the 2-form gauge potential, which is defined to be
 \be
 \Pi=2\frac{\delta {\cal L}}{\delta (\p_t A_{xr})}.
 \ee
 In the Wilson loop case, the conjugate momentum
to the gauge potential gives the charge of F1-strings. We expect
that the conjugate momentum of the 2-form gauge potential gives
the electric charge of membranes. Let us first start from the
non-chiral action. Using the nonlinear self-duality relation
(\ref{selfdual}), we have
 \be
 \Pi=\ast H,
 \ee
 where $\ast$ is respect to the induced metric of M5-brane. So, in this sense,
  the
electric charge of the Wilson surface is
 \be\label{QE2}
 Q_E=\pm\frac{R^3\sqrt{1+\k^2}}{l^3_p\k^2},
 \ee
which is different from the magnetic charge.


However, if we start from the covariant action (\ref{covariant}),
then the conjugate momentum is not so simple, it is
 \be
 \Pi^\prime=\left(\frac{\sqrt{-g}}{\sqrt{-\det(g_{mn}+i{\tilde
 H}_{mn})}}-1\right)\ast H,
 \ee
 which in this case gives
 \be
 Q^\prime_E=-(1+\frac{1}{\sqrt{1+\k^2}})Q_E.
 \ee

Therefore, we have shown that the conjugate momentum of the gauge
potential depends on the choice of the action. We are not certain
which one we should use. Fortunately, the magnetic charge is
always well-defined. It characterize the winding number of
membranes ending on the M5-brane. We will use it in our following
discussion.

 It is remarkable
that the sign in (\ref{ak}) has physical implication. From the
discussion on the charge, we know that $Q_M$ is proportional to
$-a$. So the minus sign in (\ref{ak}) means that we have membranes
ending on the M5-brane, while the plus sign in (\ref{ak})
indicates anti-membranes on M5-brane. We will see that in both
cases the soliton solution is half-supersymmetric.


It is also interesting to consider the bulk action in different
formalism. From the nonchiral action (\ref{nc}), the action of M5
brane in this case is
 \bea
 S&=&T_5\int dtdxdrd\a d\b d\g(\frac{R}{\k r})^3(\frac{R}{\k})^3\sin^2\a\sin\b
 (\frac{\k^2}{2}). \nn
 \eea

The integral over $r$ in the action shows that it is quadratically
divergent and is proportional to the area of the Wilson surface:
 \be\label{str1}
 S=\frac{N|Q_M|}{4\pi^2}TX\frac{1}{y^2_0},
 \ee
where $T$ and $X$ indicates the integral over $t$ and $x$, and
$y_0$ is a cut-off. Compared with the result from the membrane
calculation (\ref{strmembrane}) , we find that basically they
differs by a $Q_M$ factor. This fact indicates that for a Wilson
surface operator in the symmetric representation its expectation
value is $Q_M$ times the one of the fundamental representation.
Recall that $Q_M$ is the charge of the membrane and should be
identified with the rank of the representation. This is very
similar to the Wilson line case. However, there is a difference
besides $Q_M$ in the prefactor. It could be absorbed in the
cutoff. Or it indicates that these quadratic divergence should be
cancelled by appropriate counter terms, considering the BPS nature
of the configuration which will be shown in section 5.

Unlike the infinite straight Wilson line case, the action of
M5-brane is not vanishing, but is proportional to the charge. This
is not strange since our soliton solution is a self-dual one, with
both electric and magnetic charge. Even in the BPS Wilson-t' Hooft
line case, the action of D3-brane is not vanishing if not taking
into account of the boundary term\cite{ChenHe}. Our case here is
very similar. The BPS nature of our solution suggests that if we
take into account of the boundary term, the action could be
vanishing. However the boundary terms in our case seems to be
tricky. One may naively work out the conjugate momenta of $y$ and
$A_{xr}$. For the 2-form gauge potential, its conjugate momentum
has been given as above, and actually the contribution from
$\Pi^{txr}H_{txr}$ exactly cancel the bulk action.  This seems
indicate that one should only consider the boundary terms from
conjugate momentum of gauge potential. From the following
discussions on other cases, we will see that the issue is not so
simple.

One could also study the action from its covariant form
(\ref{covariant}). Since the action involves an auxiliary field,
it needs some efforts to carry it out. To simplify the
calculation, one can choose the vector $\vec{v}$ to have
nonvanishing components $v_t$ and $v_r$. With this choice, one can
check that the generalized self-dual condition (\ref{selfdual}) is
satisfied and
 \be
 {\sqrt{-\det(g_{mn}+i{\tilde
 H}_{mn})}}=\sqrt{(-g)(1+\frac{1}{2}\Tr {\tilde H}^2)}.
 \ee
 The straightforward calculation shows that the action of the solution is
 identical to the one from nonchiral action. However, one should
 note that since the conjugate momentum of the gauge potential
 from  covariant action is different from the one from nonchiral action,
 the boundary term gives the
 different contribution. It is not clear which action one should
 use to discuss the boundary terms. It turns out for the soliton solutions
 studied in this paper, the two bulk actions are the same. For the boundary terms,
 we will just  focus on the ones from the nonchiral action.


Note that for this solution, we can take M5-brane as the blow-up
of M2-brane. This is reminiscent of the D3-brane description of
the Wilson line studied in \cite{Drukker}. In the Wilson line
case, if the Wilson operator belongs to the symmetric
representation, its brane description is D3-brane, which has the
worldvolume $AdS_2\times S^2$ embedded in $AdS_5$. While if the
Wilson operator belongs to the antisymmetric representation, its
brane description is a D5-brane, which has the worldvolume
$AdS_2\times S^4$ with $AdS_2$ in $AdS_5$ and $S^4$ in
$S^5$\cite{Yamaguchi:2006D5}. In our case, we have a M5-brane
description of the infinite straight Wilson surface.  Since this
M5-brane worldvolume is completely embedded in $AdS_7$, analogue
to the Wilson line case, we may take the M5-brane solution
discussed in this section correspond to the Wilson surface
operator in the ``symmetric representation".
Intuitively, we can take the Wilson surface in the ``symmetric
representation" as the multi-wound Wilson surface. Later on, we
will see that there is another M5-brane description of the Wilson
surface in the ``antisymmetric representation", where the M5-brane
worldvolume is still a $AdS_3\times {\tilde S}^3$ but with
${\tilde S}^3$ in $S^4$.

\subsection{Spherical Wilson surface}

The spherical Wilson surface could be obtained from the straight
Wilson surface through a conformal transformation. The spherical
Wilson surface in the fundamental representation was firstly
studied in \cite{Corrado} in the context of AdS/CFT
correspondence. Unlike the straight Wilson surface, whose membrane
boundary is a two-plane, the boundary of membrane for a spherical
Wilson surface is a two-sphere. With the same philosophy, one can
get the action of the membrane\cite{Corrado}
 \be\label{M2action2}
 S=2N\left(\frac{2L^2}{\epsilon^2}-2\ln \frac{2L}{\epsilon}-1 + {\cal
 O}(\epsilon)\right)
 \ee
 where $L$ is the radius of two-sphere. It has both the quadratic
 and logarithmic divergences.

 In order to consider the M5-brane description of the
spherical Wilson surface, it is more convenient to work in the
Euclidean signature as in \cite{Drukker} and start with the
following metric of $AdS_7$:
 \be\label{metric7}
 ds^2=\frac{R^2}{y^2}(dy^2+dr_1^2+r_1^2(d\a^2+\sin^2\a
 d\b^2)+dr_2^2+r_2^2(d\g^2+\sin^2\g d\d^2).
 \ee

The Wilson surface will be placed at $r_1=L$ and $r_2=0$. Let us
change the coordinates $(r_1,r_2,y)$ to $(\rho, \eta, \theta)$ by
the following relation: \be
r_1=\frac{L\cos\eta}{\cosh\rho-\sinh\rho\cos\th},~~r_2=\frac{L\sinh\rho\sin\th}{\cosh\rho-\sinh\rho\cos\th}
~~,y=\frac{L\sin\eta}{\cosh\rho-\sinh\rho\cos\th}, \ee then we
have the $AdS_7$ metric as
 \be
 ds^2=\frac{R^2}{\sin^2\eta}\big(d\eta^2+\cos^2\eta(d\a^2+\sin^2\a
 d\b^2)+d\rho^2+\sinh^2\rho(d\th^2+\sin^2\th d\g^2+\sin^2\th
 \sin^2\g d\d^2)\big)\label{AdS7metric}
 \ee
 Here, the coordinates take the range
 $\rho\in[0,\infty),\th,\a,\g\in[0,\pi),\b,\d \in [0,2\pi),\eta
 \in [0,\pi/2)$.

To find the appropriate M5-brane that describes the blow-up of the
Wilson surface, we may take $(\rho,\a,\b,\th,\g,\d)$ as the
worldvolume coordinates of M5-brane and assume that $\eta$ be only
the function of $\rho$. Equivalently, we can think $\eta$ instead
of $\rho$ as the worldvolume coordinate. Inspired by the solution
in D3-brane of the Wilson line, we make the following ansatz
between $\eta$ and $\rho$:
 \be
 \sin \eta = \k^{-1} \sinh \rho,
 \ee
then the induced metric is
 \be
 ds^2=\frac{R^2}{\sin^2\eta}\big(\frac{1+\k^2}{1+\k^2\sin^2\eta}d\eta^2+\cos^2\eta(d\a^2+\sin^2\a
 d\b^2)\big)+R^2\k^2(d\th^2+\sin^2\th d\g^2+\sin^2\th
 \sin^2\g d\d^2).\label{indmetric2}
 \ee

We turn on the self-dual field strength on the M5-brane:
 \be
 h_3=2a\big(i(\frac{R}{\sin\eta})^3\sqrt{\frac{1+\k^2}{1+\k^2\sin^2\eta}}\cos^2\eta\sin\a
 d\eta\w d\a \w d\b
 +R^3\k^3\sin^2\th\sin\g d\th \w d\g \w d\d\big).
 \ee
Notice that due to the Euclidean signature, there is a factor $i$
in $h_{\eta\a\b}$. Similarly we can work out $k^{mn},k^2, Q$ and
open membrane metric $G^{mn}$. The field strength $H_3$ is just
 \be
 H_3=2a\big(i\frac{1}{1+a^2}(\frac{R}{\sin\eta})^3\sqrt{\frac{1+\k^2}{1+\k^2\sin^2\eta}}\cos^2\eta\sin\a
 d\eta\w d\a \w d\b
 +\frac{1}{1-a^2}R^3\k^3\sin^2\th\sin\g d\th \w d\g \w d\d\big).
 \ee

With these setups, let us check if they satisfy the equation of
motion. The components of Levi-Civita connection  of the metric
(\ref{indmetric2}) are listed in Appendix. It is straightforward
but tedious  to check that the tensor equation is satisfied. For
the scalar equation, we have
 \bea
 {\cal E}^{\u 1}_\eta=\frac{R}{\sin\eta},~~\cE^{\u
 2}_\a=\frac{R\cos\eta}{\sin\eta},~~\cE^{\u
 3}_\b=\frac{R\cos\eta\sin\a}{\sin\eta},\nn\\
 \cE^{\u 4}_\eta=\frac{\k
 R\cos\eta}{\sin\eta\sqrt{1+\k^2\sin^2\eta}},~~\cE^{\u 5}_\th=\k
 R,~~\cE^{u 6}_\g=\k R\sin\th,~~\cE^{\u 7}_\d=\k R\sin\th\sin\g.
 \eea
Here the vielbein of the metric (\ref{AdS7metric}) are
 \bea
 \hat{\th}^{\u 1}=\frac{R}{\sin\eta}d\eta,~~\hat{\th}^{\u
 2}=\frac{R\cos\eta}{\sin\eta}d\a,~~\hat{\th}^{\u
 3}=\frac{R\cos\eta\sin\a}{\sin\eta}d\b,~~\hat{\th}^{\u
 4}=\frac{R}{\sin\eta}d\rho,\nn\\
 \hat{\th}^{\u 5}=\frac{R\sinh\rho}{\sin\eta}d\th,~~
 \hat{\th}^{\u 6}=\frac{R\sinh\rho\sin\th}{\sin\eta}d\g,~~
 \hat{\th}^{\u
 7}=\frac{R\sinh\rho\sin\th\sin\g}{\sin\eta}d\d.\label{vielbein7}
 \eea
From the scalar equation, we obtain one nontrivial relation coming
from the cases when ${\u c}={\u 1}$ or ${\u 4}$:
 \be
 \frac{\k}{\sqrt{1+\k^2}}=-\frac{1-a^2}{1+a^2}.
 \ee
 This is actually the same relation (\ref{ak}) if we change $\k
 \rightarrow \frac{1}{\k}$.

The charges of the string is the same as the ones in the straight
Wilson surface case, once we take into account the difference of
the parameter $\k$ in two cases.

The non-chiral action gives us
 \be
 S_{M5}=T_5\int \ast K,
 \ee
where $K=-\frac{1+a^4}{1-a^4}$.

The Wess-Zumino part of the action is more involved. The bulk
6-form gauge potential is
 \bea
 C_6&=&(\frac{R}{y})^6r_1^2r_2^2\sin\a\sin\g dr_1\w d\a\w d\b\w
 dr_2\w d\g \w d\d \nn\\
  &=&R^6\frac{\cos^3\eta\sinh^3\rho\sin^2\th\sin\a\sin\g}{\sin^6\eta}d\rho \w
  d\a\w d\b \w d\th \w d\g\w d\d \nn\\
  &
  &-R^6\frac{\cos^2\eta\sinh^2\rho\sin^3\th\sin\a\sin\g}{\sin^5\eta(\cosh\rho-\sinh\rho\cos\th)}d\eta
  \w d\a \w d\b \w d\rho\w d\g\w d\d \nn\\
  & &+R^6\frac{\cos^2\eta\sinh^3\rho\sin^2\th\sin\a\sin\g(\sinh\rho-\cos\th\cosh\rho)}{\sin^5\eta(\cosh\rho-\sinh\rho\cos\th)}d\eta
  \w d\a \w d\b \w d\th\w d\g\w d\d\nn\\
  \eea

 The  total bulk action of M5-brane turns
out to be
 \bea
 S&=&T_5(8\pi^3
 R^6)(\frac{\k^2}{4})\left(\frac{1}{\eta^2_0}+\ln
 \eta_0\right)\nn\\
  &=&\frac{N|Q_M|}{2\pi}\left(\frac{1}{\eta^2_0}+\ln
 \eta_0\right)\label{sph1}
 \eea
 where $\eta_0$ is a cutoff near $0$. It is remarkable that from the form of the Wess-Zumino
 action there could be quartic divergence. However it turns out to be vanishing in the end. So the bulk action is
 actually of both quadratic and logarithmic divergences, with the similar structure as (\ref{M2action2}). Now $\k^2=(8\pi
 N)^{-1} Q_M$, which is very small in the large N limit.
Here we have made the conformal transformation so the radius of
the sphere does not appear in the above expression. To compare
with the existing result in the literature, we can replace
$1/\eta_0$ with $L/\eta_0$ in the above relation, where $L$ is the
radius of sphere. It is remarkable that the bulk action is linear
in the charge $Q_M$ of Wilson surface. This is consistent with the
result from the field theory calculation\cite{Corrado}. The
divergence above has two origins, one is from the conformal
anomaly which relates the straight Wilson surface to the spherical
one, the other is from the divergence in the original straight
Wilson surface. Since the boundary of Wilson surface is
1-dimensional, it will not induce any conformal anomaly.

One may wonder the boundary terms also contribute to the action.
Especially one may wonder if the contribution from gauge potential
part could cancel the above divergence exactly. Unfortunately it
is not the case anymore. After taking into account of it, the
nonchiral action is still logarithmically divergent:
 \be
 S_{M5}+\Pi^{txr}H_{txr} \sim \frac{\k^4}{4}\ln \eta_0.
 \ee
 Since $\k^2$ is very small, the above contribution is next leading
 order result. In other words, the leading order divergent term is actually
 cancelled by the boundary term.

It is remarkable that unlike the Wilson loop case, there is no
finite contribution from the integral directly, no matter if or
not we take into account of the boundary terms.

 Therefore we have a different story on
Wilson surface from Wilson line. In the Wilson line case, the
expectation of the straight Wilson line is vanishing and the one
of the circular Wilson line get the contribution from the
conformal anomaly of the boundary. In the Wilson surface case, the
expectation values of both kinds of Wilson surfaces are not
vanishing.

\section{M5-brane description of the Wilson surface in the antisymmetric representation}

In the study of the brane picture of the Wilson-loop, one knows
that for the Wilson loop in the anti-symmetric representation it
should be described by D5-brane whose worldvolume is of topology
$AdS_2\times S^4$ with $AdS_2$ being in $AdS_5$ and $S^4$ in
$S^5$\cite{Yamaguchi:2006D5}. In the case of the Wilson surface,
one may expect that there exit another M5-brane description. We
will show in this section this is true. We find that though this
M5-brane is of the same topology $AdS_3\times {\tilde S}^3$,
unlike the case we studied in the above sections, the ${\tilde
S}^3$ part is embedded in $S^4$.

\subsection{Straight Wilson surface}

Let the worldvolume coordinates of M5-branes be $\xi_i$,
$i=0,\cdots 5$ and the embedding be
 \bea
 \xi_0=t,~~ \xi_1=x, ~~\xi_2=y, \nn\\
 \xi_3=\z_2, ~~\xi_4=\z_3,~~\xi_5=\z_4,~~ \z_1=\z^0
 \eea
 where $\z_i$ are the angular coordinates of $S^4$. Here we let
 $\z_1$ be fixed at a constant $\z^0$. The induced metric is
 \bea\label{indmetric3}
 ds^2_{\mbox{ind}}=\frac{R^2}{y^2}(-dt^2+dx^2+dy^2)+\frac{R^2\sin^2\z^0}{4}(d\z_2^2
 +\sin^2\z_2d\z_3^2+\sin^2\z_2\sin^2\z_3d\z^2_4).
 \eea

In this case, we take the self-dual 3-form field strength on the
M5-brane worldvolume to be \bea
 h_3=2aR^3(\frac{1}{y^3}dt\w dx \w dy+
 \frac{\sin^3\z^0}{8}\sin^2\z_2 \sin\z_3 d\z_2\w d\z_3 \w d\z_4)
 \eea

Similar to the above cases, we can get $k^{mn}$,
$k^2=\frac{3}{2}a^4$ and  $Q=1-a^4$. The open membrane metric
$G^{mn}$ take the diagonal form:
 \bea
 G^{tt}=-G^{xx}=-G^{yy}=(1+a^2)^2(\frac{y}{R})^2,
 ~~\nn\\
 G^{22}=(1-a^2)^2\frac{4}{R^2\sin^2\z^0},~~
 G^{33}=\frac{G^{22}}{\sin^2\z_2},~~G^{44}=\frac{G^{22}}{\sin^2\z_2\sin^2\z_3},
 \eea
where $G^{ii}$ denotes $G^{\z_i\z_i}$. And the physical 3-form is
 \bea
 H_3=2aR^3(\frac{1}{(1+a^2)y^3}dt\w dx \w dy+
 \frac{\sin^3\z^0}{8(1-a^2)}\sin^2\z_2 \sin\z_3 d\z_2\w d\z_3 \w
 d\z_4),
 \eea
satisfying $dH_3=0$.

It is straightforward to check if it is possible and under what
condition if possible that the above ansatz satisfy the equations
of motion.   The tensor equation holds under the above setup. For
the scalar equation, the $AdS_3$ part is trivially satisfied. For
the ${\tilde S}^3$ part, we have
 \bea
 \cE^{\u 2}_{\z_2}=\frac{R\sin\z^0}{2}, ~~\cE^{\u 3}_{\z_3}=\frac{R\sin\z^0\sin\z_2}{2}, ~~
 \cE^{\u 4}_{\z_4}=\frac{R\sin\z^0\sin\z_2\sin\z_3}{2},
 \eea
 where we set the vielbein of $S^4$ part to be
 \bea\label{vielbein4}
 \hat{\th}^{\u 1}=\frac{R}{2}d\z_1,~~\hat{\th}^{\u
 2}=\frac{R}{2}\sin\z_1 d\z_2,~~
 \hat{\th}^{\u 3}=\frac{R}{2}\sin\z_1\sin\z_2 d\z_3,~~
 \hat{\th}^{\u 4}=\frac{R}{2}\sin\z_1\sin\z_2\sin\z_3d\z_4\nn\\
 \eea
 We list the relevant Christoffel symbol with respect to (\ref{indmetric3}) and the spin
connection with respect to (\ref{vielbein4}) in Appendix.

The nontrivial relation for the scalar equation comes from ${\u
c}={\u 1}$. Here the left hand side of the equation is not
vanishing due to the nonvanishing contribution from the spin
connection. And on the right hand side, since $H_{{\u 1}{\u 2}{\u
3}{\u 4}}=\frac{6}{R}$, it gives nonvanishing contribution. This
leads to a relation
 \be
 1-a^2=-2\frac{a\sin\z^0}{\cos\z^0}\label{az}
 \ee
 or
 \be\label{az1}
 a=\frac{\pm1+\sin\z^0}{\cos\z^0}.
 \ee
 Therefore, we have obtained another M5-brane soliton solution
 once (\ref{az1}) is satisfied. This solution is the same one
 in section 2.2 in \cite{Lunin07}, discussed in PST formalism.

 Let us calculate the charges of the Wilson surface. The magnetic
 charge is easy to obtain:
 \be\label{Qm2}
 Q_M=-\frac{R^3\sin^2\z^0\cos\z^0}{8l^3_p}.
 \ee
 For the electric charge, it is much subtler. One may define it from (\ref{QE}), which
 gives you
 \be
 Q_E=\frac{R^3\sin^3\z^0\cos\z^0}{8l^3_p}.
\ee On the other hand, one can define the electric charge from the
conjugate momentum.
 In this case, the
 conjugate momentum not only get contribution from the non-chiral
 action, but also from the Wess-Zumino part. And in the
 Wess-Zumino part of the action, there exist an ambiguity in
 defining the 3-form gauge potential. We make the following choice
 \be
 C_3=-\frac{3}{8}R^3(-\cos\z^0+\frac{1}{3}\cos^3\z^0)\sin^2\z_2\sin\z_3.
 d\z_2\w d\z_3 \w d\z_4
 \ee
Then the electric charge is
 \be\label{Qe2}
 Q_E=\frac{R^3}{16l^3_p}\cos\z^0\big(\sin^3\z^0+3-\cos^2\z^0\big).
 \ee

Similar to the case in section 3, the sign in (\ref{az1}) is
physical. The situation here is a little subtler. No matter which
sign we take, we always get the same formulae on magnetic and
electric charges. In other words, what kind of membrane the
M5-brane feels depends on $\cos\z^0$ rather than the sign in
(\ref{az1}). Nevertheless, we will show that the different choice
of sign indicates the different supersymmetries the soliton
solution keeps.


Note that once we turn off the 3-form field on M5-brane, there
still exists an M5-brane solution, which reside at $\z^0=\pi/2$.
This means that the M5-brane without flux could be embedded in the
background without instability.

 The bulk action is
 \bea
 S&=&T_5 \int (\ast K-2 {\u C}_3\w H_3) \nn\\
  &=&\frac{T_5}{8}4\pi^2R^6TX\frac{1}{2y_0^2}\nn\\
  &=&\frac{N^2}{2\pi}TX\frac{1}{y_0^2},\label{str2}
  \eea
which is quadratically divergent, and similar to
(\ref{strmembrane}). Now since $Q_M\sim N$, the action is still
proportional to $NQ_M$. Similarly, if we try to take the
contribution from boundary terms coming from the conjugate
momentum of gauge potential into account, we have
 \bea
 S\sim \sin^2\z^0\frac{1}{y_0^2}.
 \eea

\subsection{Spherical Wilson surface}

For the spherical Wilson surface, we have to do a conformal
transformation of the above one. The embedding of ${\tilde S}^3$
in $S^4$ is the same as before. For the $AdS_3$ part, it is
somehow different. The metric of Euclideanized $AdS_7$ take the
form (\ref{metric7}). Let us assume that the spherical Wilson
surface satisfy $r_1^2+y^2=L^2$, namely a sphere $S^2$ with radius
$L$, and later on we will check such kind of embedding satisfies
the equations of motion. Let
 \be
 y=L\cos\d, ~~r_1=L\sin\d,
 \ee
then the $AdS_3$ part of the induced metric of M5-brane is
 \be\label{indmetric4}
 ds^2_{\mbox{ind}}=\frac{R^2}{\cos^2\d}(d\d^2+\sin^2\d(d\a^2+\sin^2\a
 d\b^2))
 \ee

The self-dual 3-form field strength on the M5-brane worldvolume
could be set to  \bea
 h_3=2aR^3(i\frac{\sin^2\d\sin\a}{\cos^3\d}d\d\w d\a \w d\b+
 \frac{\sin^3\z^0}{8}\sin^2\z_2 \sin\z_3 d\z_2\w d\z_3 \w d\z_4).
 \eea
From it, we can calculate the other quantities as before. The only
differences from the straight case are
 \be G^{\d\d}=(1+a^2)^2(\frac{\cos\d}{R})^2,~~
 G^{\a\a}=\frac{G^{\d\d}}{\sin^2\d},~~G^{\b\b}=\frac{G^{\d\d}}{\sin^2\d\sin^2\a}.
 \ee
 The physical 3-form field strength is now
 \bea
 H_3=2aR^3(i\frac{1}{1+a^2}\frac{\sin^2\d\sin\a}{\cos^3\d}d\d\w d\a \w d\b+
 \frac{1}{1-a^2}\frac{\sin^3\z^0}{8}\sin^2\z_2 \sin\z_3 d\z_2\w d\z_3 \w
 d\z_4).\nn\\
 \eea

 For the scalar equation, the discussion on ${\tilde S}^3$ part
does not change and we find the same relation as (\ref{az}). We
needs to check if the Euclideanized $AdS_3$ part does not give
anything nontrivial. This could be checked explicitly. Now we have
 \be
 {\cal E}^{\u 1}_\d=-\frac{R\sin\d}{\cos\d},~~{\cal E}^{\u
 2}_\d=R,~~{\cal E}^{\u 3}_\a=\frac{R\sin\d}{\cos\d},~~
 {\cal E}^{\u 4}_\b=\frac{R\sin\d}{\cos\d}\sin\a,
 \ee
where we have set the relevant vierbeins to be
 \be
 \hat{\th}^{\u 1}=\frac{R}{y}dy,~~\hat{\th}^{\u 2}=\frac{R}{y}dr_1,~~
 \hat{\th}^{\u 3}=\frac{R}{y}r_1d\a,~~\hat{\th}^{\u
 4}=\frac{R}{y}r_1\sin\a d\b.\label{vierbein4}
 \ee
 Here we abuse the indices which we wish would not bring any
 confusion to the reader. From the embedding, it is not obvious
 that $\nabla_m{\cal E}_m^{\u c}=0$. However the explicit
 calculation shows that this is indeed true. The relevant
 Levi-Civita connection and the spin connection are put into
 Appendix.

Unlike the case discussed in section 3, the conformal
transformation from straight surface to sphere is somehow trivial.
Therefore the charges of the membrane on M5-brane is the same as
(\ref{Qm2}, \ref{Qe2}). The bulk action reads
 \bea
 S&=&2\pi^3R^6T_5\int^{\frac{\pi}{2}}_0 \frac{\sin^2\d}{\cos^3\d}
 \nn \\
  &=&\frac{\pi^3}{2}R^6T_5
  (\frac{2}{\epsilon^2}-\ln\frac{2}{\epsilon})\nn\\
  &=&N^2(\frac{2}{\epsilon^2}-\ln\frac{2}{\epsilon}),\label{sph2}
  \eea
where $\epsilon$ is the cutoff near $\d=\frac{\pi}{2}$. The terms
in the bracket of (\ref{sph2}) looks familiar. Actually, in the
above discussion we have chose the angular coordinates so that the
radius of the sphere does not show up. It is easily to recover it
by replace $1/\epsilon$ with $L/\epsilon$. Then we have the
similar divergent terms as (\ref{M2action2}). The significant
difference is that the M5-brane action is proportional to the
membrane charge $Q_M$.

\section{Supersymmetry analysis}

Let us check if the above solutions are supersymmetric. Firstly we
need to work out the Killing spinor of the bulk background. From
the discussions above, we notice that all the solution has a
global symmetry $SO(2,2)\times SO(4)\times SO(4)$. In order to
make the analysis simpler, we first rewrite $AdS_7\times S^4$
metric in form of $AdS_3\times S^3 \times {\tilde S}^3$ fibred
over two-dimensional base space: \be\label{fibremetric}
ds^2=R^2\left(\cosh^2\rho\,ds^2_{AdS_3}+\sinh^2\rho\,d\Omega^2_3+d\rho^2\right)+{R^2\over
4}\left(d\z_1^2+\sin^2\z_1\,d\tilde\Omega^2_3\right),\ee In the
new fibred coordinates, the $4$-form flux can be written as: \be
H_4={6\over R}e^7\wedge e^8 \wedge e^9 \wedge e^{10}. \ee Here we
use $ds^2_{AdS_3}, d\Omega^2_3, d\tilde\Omega^2_3$ to denote the
metric of unit $AdS_3$ and two unit $S^3$'s, respectively. The
$e^M$'s, $M=0, \cdots, 10$ are the vielbein of this metric.

We use $\Gamma^M$ to denote the 11-dimensional Gamma matrices.
They can be written as the following form\cite{Yamaguchi:2006te}:
\begin{eqnarray}
\Gamma^0&=&\gamma_8\otimes\check{\sigma}^0\otimes 1\otimes
1\otimes \sigma_1,
~~~\Gamma^1=\gamma_8\otimes\check{\sigma}^1\otimes
1\otimes 1\otimes \sigma_1, \nn\\
\Gamma^2&=&\gamma_8\otimes\check{\sigma}^2\otimes 1\otimes
1\otimes \sigma_1, ~~~\Gamma^3=\gamma_8\otimes1\otimes
\hat{\sigma}^3\otimes 1\otimes \sigma_2,\nonumber \\
\Gamma^4&=&\gamma_8\otimes1\otimes \hat{\sigma}^4\otimes 1\otimes
\sigma_2, ~~~\Gamma^5=\gamma_8\otimes1\otimes
\hat{\sigma}^5\otimes 1\otimes \sigma_2, \nonumber\\
\Gamma^6&=&\gamma^6\otimes1\otimes 1\otimes 1\otimes 1, \,\,\,\,\,
~~~\Gamma^7=\gamma^7\otimes1\otimes 1\otimes 1\otimes
1, \nonumber \\
\Gamma^8&=&\gamma_8\otimes1\otimes 1\otimes \tilde\sigma^8\otimes
\sigma_3, ~~~\Gamma^9=\gamma_8\otimes1\otimes 1\otimes
\tilde\sigma^9\otimes \sigma_3, \nonumber \\
\Gamma^{10}&=&\gamma_8\otimes1\otimes 1\otimes
\tilde\sigma^{10}\otimes \sigma_3. \label{eqgamma}
\end{eqnarray}
Here $(\gamma^6, \gamma^7, \gamma_8), (\check{\sigma}^1,
\check{\sigma}^2, \check{\sigma}^3), (\hat{\sigma}^3,
\hat{\sigma}^4, \hat{\sigma}^5), (\tilde\sigma^8, \tilde\sigma^9,
\tilde\sigma^{10}), (\sigma_1, \sigma_2, \sigma_3)$ are five sets
of Pauli matrices and $\check{\sigma}^0=i\check{\sigma}^3$.

The first step is to find the Killing spinor in $AdS_7\times S^4$
with the above fibred coordinates. In order to do so, we need to
use the following Killing spinors of the unit $AdS_3$ and the two
unit $S^3$'s: \be
\mathring{\nabla}_p\check{\chi}^I_{a^\prime}={i\over2}{a^\prime}\check{\sigma}_p\check{\chi}^I_{a^\prime},\,(p=0,
1, 2, {a^\prime}=\pm 1, I=1, 2), \ee \be
\mathring{\nabla}_p\hat{\chi}^J_{b^\prime}={1\over2}{b^\prime}\hat{\sigma}_p\hat{\chi}^J_{b^\prime},\,(p=3,
4, 5, {b^\prime}=\pm 1, J=1, 2), \ee \be
\mathring{\nabla}_p\chi^K_{c^\prime}={i\over2}{c^\prime}\sigma_p\chi^K_{c^\prime},\,(p=8,
9, 10, {c^\prime}=\pm 1, K=1, 2), \ee and decompose the
$11$-dimensional spinor $\xi$ as
\begin{equation}
\xi=\sum_{{a^\prime}{b^\prime}{c^\prime}IJK}
\epsilon_{{a^\prime}{b^\prime}{c^\prime}IJK}\check{\chi}^I_{a^\prime}\otimes\hat{\chi}^J_{b^\prime}\otimes\chi^K_{c^\prime},
\end{equation}
where each $\epsilon_{{a^\prime}{b^\prime}{c^\prime}IJK}$ is a
pair of $2$-dimensional spinors with $\g^6, \g^7, \g_8, \sigma_1,
\sigma_2, \sigma_3$ acting on it. In another word, each
$\epsilon_{a^\prime b^\prime c^\prime IJK}$ belongs to the tensor
product of the space of the $2$-dimensional spinor and ${\bf
C}^2$, and $\g^6, \g^7, \g_8$ act on the space of the
$2$-dimensional spinor, while $\s_1, \s_2, \s_3$ act on ${\bf
C}^2$.

From the Killing spinor equation and the above decomposition,  we
obtain the following supersymmetric conditions:
 \be
({a^\prime}\sigma_1+i\sinh\rho\gamma^7-i\cosh\rho\gamma^7\sigma_3)\epsilon=0,
\label{eqsusy1}\ee \be
({b^\prime}\sigma_2+\cosh\rho\gamma^7-\sinh\rho\gamma^7\sigma_3)\epsilon=0,
\label{eqsusy2}\ee \be
({c^\prime}\sigma_3-\cos\z_1\gamma^6+\sin\z_1\gamma^7\sigma_3)\epsilon=0,
\label{eqsusy3}\ee
\begin{equation}
{\partial\epsilon\over\partial\rho}={1\over2}\sigma_3\,\epsilon,~~~
{\partial\epsilon\over\partial\z_1}={i\over2}\gamma_8\sigma_3\,\epsilon.
\label{eqsusyf}
\end{equation}

Now we begin to solve these equations. Using eq.~(\ref{eqsusyf}),
we find that $\epsilon$ can be written as
\begin{equation}
\epsilon=\exp\left({1\over2}\sigma_3\rho+{i\over2}\gamma_8\sigma_3\z_1\right)\zeta.
\end{equation}
Here $\zeta$ is a pair of constant spinors. Then
eqs.~(\ref{eqsusy1}-\ref{eqsusy3}) lead to the following
projective conditions:
\begin{equation}
{a^\prime}\g^7\sigma_2\zeta=\zeta,~~~
{b^\prime}\g^7\sigma_2\zeta=-\zeta,~~~
{c^\prime}\g^6\sigma_3\zeta=\zeta,\label{killing}
\end{equation}
Since for each $a^\prime, b^\prime, c^\prime, I, J, K$,
$\epsilon_{a^\prime b^\prime c^\prime IJK}$ is a pair of two
dimensional spinors and $a^\prime, b^\prime, c^\prime=\pm1, I, J,
K=1, 2$, totally there are $2^8$ components. After imposing $3$
projection conditions each of which project out half of the
components, we have $2^5$ complex components. Finally imposing
Majorana condition leaves $2^5$ real components.

Now let us introduce the Gamma matrix $\Gamma_{M5}$, which is
determined by the M5-brane worldvolume and the flux on it\cite{Sezgin97}:
 \begin{equation}
 \Gamma_{M5}=\frac{1}{6!\sqrt{-g}}\epsilon^{j_1\cdots j_6}[\Gamma_{<j_1\cdots
 j_6>}+40\Gamma_{<j_1j_2j_3>}h_{j_4j_5j_6}].
 \end{equation}
Here $g$ is the determinant of the induced worldvolume metric
component, $h_{j_4j_5j_6}$ is the self-dual 3-form on the
M5-brane. And $\Gamma_{<j_1\cdots j_n>}$ is defined as
 \begin{equation}
 \Gamma_{<j_1,\cdots j_n>}={\cal E}^{\underline a_1}_{j_1}\cdots{\cal E}^{\underline
 a_n}_{j_n}\Gamma_{{\underline a_1}\cdots {\underline a_n}},
 \end{equation}
 where $\Gamma_{{\underline a_1}\cdots {\underline a_n}}$ is the product of the Gamma matrices
 in the frame.

The kappa symmetry projection condition is
 \begin{equation}\label{SUSY}
 \Gamma_{M5}\xi = \xi.
 \end{equation}
 The amount of unbroken supersymmetry is determined by the
 solution of above equation.

For the straight Wilson surface case, the metric of the $M5$-brane
and the self-dual $3$-form flux on it can be written in the fibred
coordinates as \be ds^2=R^2(\cosh^2\rho_k
ds^2_{AdS_3}+\sinh^2\rho_k d\Omega^2_3),\ee \bea h_3&=&{a\over
2}(e^0\wedge e^1 \wedge e^2+e^3 \wedge e^4
\wedge e^5)\nn\\
&=&{a\over 2}R^3\left(\cosh^3\rho_k\cosh\tilde\rho\sinh\tilde\rho
d\tau \wedge d\tilde\rho\wedge
d\tilde\theta+\sinh^3\rho_k\sin^2\alpha\sin\beta d\alpha\wedge
d\beta \wedge d\gamma\right)\nn\\ \eea

From this, after some short calculations, one get \be
\G_{M5}=-(\G_{01\cdots5}+a(\G_{012}-\G_{345}))\ee

Using the above representation of $\G^\mu$, one can find that the
condition (\ref{SUSY})  is equivalent to \be
-\sigma_3\,\epsilon+a\g_8\sigma_1\,\epsilon+ia\g_8\sigma_2\,\epsilon=\epsilon.
\ee

When $\z_1=0$, we have
$\epsilon=\exp({1\over2}\sigma_3\rho_k)\zeta$, then the above
equation is equivalent to \be
-\sigma_3\zeta-\zeta+ae^{-\rho_k}\g_8\sigma_1\zeta+iae^{-\rho_k}\g_8\sigma_2\zeta=0.
\ee

From this, we can obtain the following supersymmetry condition for
$a, \rho_k$ and $\zeta$: \be a=\pm
e^{\rho_k},\,~~\pm\g_8\s_1\zeta=\zeta.\ee The projection
conditions on $\zeta$ here are compatible with the projection
conditions in eqs.~(\ref{killing}) for the Killing spinors. So the
supersymmetry conditions are satisfied by half of the components
of the Killing spinors. In another word, our solution is half-BPS.
 In the case of $\z_1=\pi$, we can similarly obtain the following supersymmetry conditions:
 \be  a=\pm
e^{\rho_k},\,~~\mp\g_8\s_1\zeta=\zeta. \ee

Let us set $\sinh\rho_k=\frac{1}{\k}$ to recover the induced
metric (\ref{indmetric01}) from fibred metric (\ref{fibremetric}).
Then the relation (\ref{ak}) is exactly the relation $a=\pm
e^{\rho_k}$. This shows that our M5-brane soliton solution
corresponding to the straight Wilson surface is half-BPS. For the
spherical solution, we get the same conclusion.

For the case of $\tilde S^3$, we can similarly obtain, \be
\G_{M5}=-(\G_{01289(10)}+a(\G_{012}-\G_{89(10)})). \ee Using
eq.~(\ref{eqgamma}), we find that the relation (\ref{SUSY}) is
equivalent to \be
\sigma_2\,\epsilon+a\g_8\sigma_1\,\epsilon+ia\g_8\sigma_3\,\epsilon=\epsilon
\ee
for the straight Wilson surface.
Now, we have $\rho=0$, then
$\epsilon=\exp({i\over2}\g_8\sigma_3\z^0)\zeta$, we can find that
the above equation is equivalent to \bea (\cos{\z^0\over
2}+a\sin{\z^0\over2})\sigma_2\zeta&+&(a\cos{\z^0\over 2}-\sin{\z^0
\over 2 })\g_8\sigma_1\zeta\nn\\-(\cos{\z^0\over2}+a\sin{\z^0
\over 2})\zeta&-&i(\sin{\z^0\over 2}-a\cos{\z^0 \over
2})\g_8\sigma_3\zeta=0 \eea

This gives us the following supersymmetry conditions on $a, \z^0$
and $\zeta$, \be a={\pm1+\sin\z^0\over\cos\z^0},\,~~\pm
\g_8\s_1\zeta=\zeta.\ee The first relation is exactly (\ref{az1}).
As in the previous case, the projection condition here are also
compatible with the projection conditions eqs.~(\ref{killing}).
Then we have shown that our solution in this case is half-BPS as
well. The discussion on the spherical Wilson surface is similar.

From the above discussion, we come to the conclusion that all our
solutions are half-BPS.

\section{Conclusion and discussion}

In this paper, we investigated the M5-brane soliton solutions in
$AdS_7 \times S^4$ background.  Starting from the covariant
equations of motion of M5-brane, we found two classes of
solutions, both having $AdS_3\times S^3$ topology. The $AdS_3$
part is always in $AdS_7$ but $S^3$ could be in $AdS_7$ or $S^4$.
The two different configurations give the description of the
Wilson surface operators in the symmetric and the anti-symmetric
representation respectively.  We discussed the properties of these
solutions and their implications to the Wilson surface operators
from AdS/CFT correspondence. Unfortunately due to the shortage of
the discussion on the Wilson surface operators in six-dimensional
(2,0)-theory side, we were not able to make comparison more
precisely.

From the dictionary of AdS/CFT correspondence, the exponential of
bulk M5-brane action with boundary terms could give the
expectation values of the surface operators. We are not certain of
the boundary terms, which involves the conjugate momenta of the
gauge potential and coordinate. Nevertheless, there are a few
remarkable points on the bulk actions. Firstly for the straight
Wilson surface operators, the bulk actions are quadratically
divergent, and for the spherical ones, the bulk actions are both
quadratically and logarithmically divergent. These two cases are
related to each other by conformal transformation, being in
consistence with argument from conformal anomaly\cite{Witten99}.
Secondly, compared to the result on the Wilson surface operators
in the fundamental representation from the membrane approach, the
bulk M5-brane action is $Q_M$ times the membrane action up to a
numerical factor. The $Q_M$ characterizes the charges carried by
the membrane and also the rank of the representation. This fact
implies that whatever the representation the Wilson surface
operators are in, the possible M5-brane action should have the
same structure. Namely the actions take the similar form as
(\ref{str1},\ref{str2}) for the straight surfaces and
(\ref{sph1},\ref{sph2}) for the spherical surfaces, being of the
divergent terms times the rank of the representation. Thirdly, the
fact that the solutions we found are all supersymmetric indicates
that the Wilson surface operators are supersymmetric too. This
implies that their expectation values should be exactly one since
the bulk action should be vanishing after taking into account the
appropriate boundary terms. Furthermore, since there is no
conformal anomaly from boundary terms in our cases, the
implication should make sense both in the straight and the
spherical cases.

Our solutions are the examples of M5-brane self-dual string
soliton solutions in curved spacetime. These solutions have been
discussed in \cite{Lunin07} from another approach. To our
knowledge, the string soliton solutions in curved spacetime have
not been studied carefully in the literature.  The study of these
soliton solution would be quite valuable and open a new window to
the study of M5-brane physics and M-theory. To find more string
soliton solution in curved spacetime and study their properties is
an interesting question.

On the other hand, it would be very nice to understand the string
soliton configurations from the dynamics of nonabelian membranes.
In the Wilson loop case, there exist a dielectric description of
F1(D1) blowing up to higher dimensional
D-brane\cite{Rodriguez2006}. One may wonder if the same story is
true here. However, we have no good understanding of nonabelian
membrane action. In \cite{Basu}, a generalized Nahm equation has
been proposed and the funnel solution from membrane has been
constructed. But it is still an open issue how to construct the
nonabelian membrane action, even in the flat
spacetime\cite{Baggar}. In the case at hand, we need to know the
nonabelian action in curved spacetime with background flux. We
expect that some kind of Myers effect\cite{Myers:1999} exists in
M-theory. This is a very important question.

The six-dimensional (2,0) superconformal field theory is very
nontrivial. Some people have proposed the DLCQ matrix description
of the theory\cite{Seiberg97}. It would be nice to see if this
description could address the Wilson surface operators issue. This
may help us to make the AdS/CFT dictionary in this case more
precise.

The string soliton solutions constructed in this paper are
half-BPS. It would be interesting to find other string soliton
solutions with less supersymmetry. These string soliton solutions
will correspond to the membranes ending on M5-branes in
$AdS_7\times S^4$ background. For the discussion on such soliton
solutions in flat spacetime, see \cite{Yee07}.

There are several subtleties in our discussion. We are not
satisfied with the boundary terms we discussed. The main trouble
comes from the ambiguity in choosing the action. Unlike the DBI
action for D-brane in string theory, there is no well-accepted
action for M5-brane. The different action may lead to different
conjugate momenta and different boundary terms. Moreover, we are
not sure if the naive application of the prescription found in the
Wilson loop case is legal. Anyhow, the 3-form field in the
M5-brane worldvolume is quite special. We have quite poor
knowledge on it. Nevertheless, the action of the string soliton
solutions did catch the essential properties of the multi-wound
Wilson surface and the multi-Wilson surfaces.  It would be very
interesting to have a field theory calculation of the expectation
values of the Wilson surfaces.

The surface operators could also be an important order parameter
in four-dimensional gauge field theory. It has been used to study
the geometric Langlands programme with
ramification\cite{Witten06}. The bubbling geometry picture of the
surface operators in ${\cal N}=4$ SYM has been proposed in
\cite{Gomis07}. It would be interesting to illuminate the
relations between the surface operators in four-dimensional SYM
and the Wilson surface operators in six-dimensional (2,0)-theory.

\section*{Acknowledgments}
The work was partially supported by NSFC Grant No.
10405028,10535060, NKBRPC (No. 2006CB805905) and the Key Grant
Project of Chinese Ministry of Education (NO. 305001). BC would
like to thank ICTP for its hospitality, where the project was
finished.  The work of JW is supported in part by the European Community's Human Potential
Programme under contract MRTN- CT-2004-005104 `Constituents, fundamental forces and symmetries of the universe' as a postdoc of the node of Padova.

\section{Appendix: Various connections}

  In this appendix, we list various connections appeared in our
  calculation. For the induced metric (\ref{indmetric}), its Christoffel symbol
  has nonvanishing components:
  \bea
  \G^t_{rt}&=&\G^x_{xr}=-\frac{f^{\pr}}{f} \nn \\
  \G^r_{tt}&=&-\G^r_{xx}=-\frac{f^{\pr}}{(1+f^{\pr 2})f},\nn\\
  \G^r_{rr}&=&-\frac{f^{\pr}}{f}+f^\prime \frac{ f^{\prime\prime}}{1+f^{\pr 2}} \nn \\
  \G^r_{\a\a}&=&\frac{1}{1+f^{\pr 2}}(-r+\frac{f^{\pr}}{f}r^2)\nn\\
  \G^r_{\b\b}&=&\frac{1}{1+f^{\pr 2}}(-r+\frac{f^{\pr}}{f}r^2)\sin^2\a\nn\\
  \G^r_{\g\g}&=&\frac{1}{1+f^{\pr 2}}(-r+\frac{f^{\pr}}{f}r^2)\sin^2\a\sin^2\b \nn\\
   \G^\a_{r\a}&=&\G^\b_{r\b}=\G^\g_{r\g}=(\frac{1}{r}-\frac{f^{\pr}}{f})\nn\\
  \G^\a_{\b\b}&=&-\sin\a\cos\a, \nn\\
   \G^\a_{\g\g}&=&-\sin^2\b\sin\a\cos\a \nn\\
  \G^\b_{\a\b}&=&\G^{\g}_{\a\g}=\frac{\cos\a}{\sin\a},\nn\\
  \G^\b_{\g\g}&=&-\sin\b\cos\b \nn\\
   \G^\g_{\b\g}&=&\frac{\cos \b}{\sin\b}
  \eea
When $f=\k r$, some of the components above are vanishing.

  For the $AdS_7$ spacetime, its nonvanishing independent components of spin connection
   are
  \bea
  \o^{\underline 1}_{{\underline 0}{\underline 0}}&=&-\frac{1}{R},
  ~~~\o^{\underline 1}_{{\underline i}{\underline
  i}}=\frac{1}{R}, \hspace{3ex}\mbox{for $i\neq 0,1$}\nn\\
  \o^{\underline 3}_{\underline ii}&=&-\frac{y}{Rr},\hspace{3ex}\mbox{for $i=4,5,6$}\nn\\
  \o^{\underline 4}_{{\underline i}{\underline
  i}}&=&-\frac{y\cos\a}{Rr\sin\a},\hspace{3ex}\mbox{for $i=5,6$}\nn\\
  \o^{\underline 5}_{{\underline 6}{\underline
  6}}&=&-\frac{y\cos\b}{Rr\sin\a\sin\b}
  \eea

For the metric (\ref{indmetric2}), its Levi-Civita connection has
components:
 \bea
 \G^\a_{\eta \a}=\G^\b_{\eta
 \b}=-\frac{1}{\sin\eta\cos\eta},~~\G^\a_{\b\b}=-\sin\a\cos\a,~~\G^\b_{\a\b}=\frac{\cos\a}{\sin\a}\nn\\
 \G^\eta_{\eta\eta}=-\frac{\cos\eta(1+2\k^2\sin^2\eta)}{\sin\eta(1+\k^2\sin^2\eta)},~~
 \G^\eta_{\a\a}=\frac{(1+\k^2\sin^2\eta)\cos\eta}{(1+\k^2)\sin\eta},~~\nn\\
 \G^\eta_{\b\b}=\frac{(1+\k^2\sin^2\eta)\cos\eta\sin^2\a}{(1+\k^2)\sin\eta}.
 \eea
The independent nonvanishing components of the spin connections
with respect to the vielbeins (\ref{vielbein7}) are
 \bea
 \o^{\u 1}_{{\u 2}{\u 2}}=\o^{\u 1}_{{\u 3}{\u 3}}=\frac{1}{R\cos\eta},~~\o^{\u
 2}_{{\u 3}{\u 3}}=-\frac{\sin\eta\cos\a}{R\cos\eta\sin\a},\nn\\
 \o^{\u 1}_{{\u 4}{\u 4}}=\o^{\u 1}_{{\u 5}{\u 5}}=\o^{\u 1}_{{\u 6}{\u 6}}=\o^{\u 1}_{{\u
 7}{\u 7}}=\frac{\cos\eta}{R},\nn\\
 \o^{\u 4}_{{\u
 5}{\u 5}}=\o^{\u 4}_{{\u 6}{\u 6}}=\o^{\u 4}_{{\u 7}{\u 7}}=-\frac{\cosh\rho\sin\eta}{R\sinh\rho},~~\nn\\
 \o^{\u 5}_{{\u 6}{\u 6}}=\o^{\u 5}_{{\u
 7}{\u 7}}=-\frac{\cos\th\sin\eta}{R\sin\th\sinh\rho},~~\o^{\u 6}_{{\u
 7}{\u 7}}=-\frac{\cos\g\sin\eta}{R\sinh\rho\sin\th\sin\g}
 \eea

For the metric (\ref{indmetric3}), its Levi-Civita connection has
components:
 \bea
 \G^2_{33}=-\sin\z_2\cos\z_2,~~\G^2_{44}=-\sin\z_2\cos\z_2\sin^2\z_3,\nn\\
 ~~\G^3_{44}=-\sin\z_3\cos\z_3,~~
 \G^3_{23}=\G^4_{24}=\frac{\cos\z_2}{\sin\z_2},~~\G^4_{34}=\frac{\cos\z_3}{\sin\z_3}.
 \eea
 where the index $i$ indicates $\z_i$.
And the nonvanishing components of the spin connection of $S^4$
with respect to vielbeins (\ref{vielbein4}) are
 \bea
 \o^{\u 1}_{{\u 2}{\u 2}}=\o^{\u 1}_{{\u 3}{\u 3}}=\o^{\u 1}_{{\u 4}{\u
 4}}=-\frac{2\cos\z_1}{R\sin\z_1},\\
 \o^{\u 2}_{{\u 3}{\u 3}}=\o^{\u 2}_{{\u 4}{\u
 4}}=-\frac{2\cos\z_2}{R\sin\z_1\sin\z_2},~~\o^{\u 3}_{{\u 4}{\u
 4}}=-\frac{2\cos\z_3}{R\sin\z_1\sin\z_2\sin\z_3}
 \eea

For the metric (\ref{indmetric4}), its nonvanishing Levi-Civita
connection components are of the form
 \bea
 \G^\d_{\d\d}=-\G^\d_{\a\a}=\frac{\sin\d}{\cos\d},~~\G^\d_{\b\b}=-\frac{\sin\d}{\cos\d}\sin^2\a\nn\\
 \G^\a_{\d\a}=\G^\b_{\d\b}=\frac{1}{\sin\d\cos\d},~~\G^\a_{\b\b}=-\sin\a\cos\a,~~\G^\b_{\a\b}=\frac{\cos\a}{\sin\a}
 \eea
 The spin connection with respect to the vielbein
 (\ref{vierbein4}) has the components:
 \bea
 \o^{\u 1}_{{\u i}{\u i}}=\frac{1}{R},~~ (\mbox{for $i=2,3,4$}),~~ \o^{\u 2}_{{\u 3}{\u
 3}}=-\frac{y}{Rr_1},\nn\\
\o^{\u 2}_{{\u 4}{\u 4}}=-\frac{y}{Rr_1},~~\o^{\u 3}_{{\u 4}{\u
 4}}=-\frac{y\cos\a}{Rr_1\sin\a}.
 \eea

\bigskip


\end{document}